%% file: main.tex
\documentclass[sigconf]{acmart}
\AtBeginDocument{%
  }

\input{macros}
\usepackage{algorithm}
\usepackage{algpseudocode}
\usepackage{amsmath}
\usepackage{multirow}

\copyrightyear{2026}
\acmYear{2026}
\setcopyright{cc}
\setcctype{by}
\acmConference[ICSE-SEIP '26]{2026 IEEE/ACM 48th International Conference on Software Engineering}{April 12--18, 2026}{Rio de Janeiro, Brazil}
\acmBooktitle{2026 IEEE/ACM 48th International Conference on Software Engineering (ICSE-SEIP '26), April 12--18, 2026, Rio de Janeiro, Brazil}
\acmPrice{}
\acmDOI{10.1145/3786583.3786859}
\acmISBN{979-8-4007-2426-8/2026/04}

\begin{document}

\title[Smart Paste: Automatically Fixing Copy/Paste for Google Developers]{Smart Paste: Automatically Fixing Copy/Paste\texorpdfstring{\\}{ }for Google Developers}

\author{Vincent Nguyen}
\affiliation{
  \institution{Google}
  \city{New York}
  \country{USA}
}
\email{vincentdnguyen@google.com}

\author{Guilherme Herzog}
\affiliation{
  \institution{Google}
  \city{Munich}
  \country{Germany}
}
\email{guiherzog@google.com}

\author{José Cambronero}
\affiliation{
  \institution{Google}
  \city{Atlanta}
  \country{USA}
}
\email{jcambronero@google.com}

\author{Marcus Revaj}
\affiliation{
  \institution{Google}
  \city{Munich}
  \country{Germany}
}
\email{marcusrevaj@google.com}

\author{Aditya Kini}
\affiliation{
  \institution{Google}
  \city{Sunnyvale}
  \country{USA}
}
\email{akini@google.com}

\author{Alexander Frömmgen}
\affiliation{
  \institution{Google}
  \city{Munich}
  \country{Germany}
}
\email{froemmgen@google.com}

\author{Maxim Tabachnyk}
\affiliation{
  \institution{Google}
  \city{Munich}
  \country{Germany}
}
\email{tabachnyk@google.com}

\renewcommand{\shortauthors}{Nguyen et al.}

\begin{abstract}
\input{abstract}
\end{abstract}

\maketitle

\input{intro}

\input{problem}

\input{data}
\input{training}
\input{ux}
\input{results}
\input{discussion}
\input{threats}
\input{relatedwork}

\input{futurework}
\input{conclusion}
\input{acknowledgement}

\bibliographystyle{ACM-Reference-Format}
\bibliography{references}
\end{document}

%% file: macros.tex
\usepackage{listings}
\usepackage{xcolor}

\newcommand{\boldlead}[1]{\vspace{0.3em}\noindent{\textbf{#1}}}
\newcommand{\leadbold}[1]{\boldlead{#1}}

\definecolor{codegray}{gray}{0.95}

%% file: abstract.tex
Manually editing pasted code is a long-standing developer pain point. In internal software development at Google, we observe that code is pasted 4 times more often than it is manually typed. These paste actions frequently require follow-up edits, ranging from simple reformatting and renaming to more complex style adjustments and cross-language translations. Prior work has shown deep learning can be used to predict these edits. In this work, we show how to iteratively develop and scale Smart Paste, an IDE feature for post-paste edit suggestions, to Google’s development environment. This experience can serve as a guide for AI practitioners on a holistic approach to feature development, covering user experience, system integration, and model capabilities. Since deployment, Smart Paste has had overwhelmingly positive feedback with a 45\% acceptance rate. At Google's enterprise scale, these accepted suggestions account substantially for over 1\% of all code written company-wide.

%% file: intro.tex
\section{Introduction}
Copy-pasting is ubiquitous in developer workflows, allowing rapid code duplication. However, this often necessitates a repetitive series of manual edits to adapt the pasted code to its new syntactic and semantic context. This manual fixing process is an involved and error prone task~\cite{ray2013detecting}, encompassing a wide variety of necessary adjustments. These can range from standardized reformatting to conform to language-specific style guidelines, to minor variable renaming, or incrementing enumerated types. More complex pastes, such as integrating code from different sources or those that include new method calls, can demand familiarity with cross-language translations or library-specific APIs. These edits are tedious to perform, and misidentifying or incorrectly addressing them can easily lead to errors. 
Existing developer features in modern IDEs offer only partial solutions to the paste fixing problem:

\begin{itemize}
    \item Code completion~\cite{bruch2009learning, svyatkovskiy2019pythia, raychev2014code} can expedite fixes by offering predictive suggestions which can be accepted by tab completion to reduce keystrokes. However, this can only address incremental additions at the end of a paste region.
    
\item Code formatters or linters~\cite{tomasdottir2017and} can address a subset of issues like style and dependencies, but tasks such as context-aware renaming or logic changes remain out of scope. 

\item Chatbots~\cite{team2023gemini, achiam2023gpt}, or other agentic approaches to software development~\cite{yang2024swe}, can make many types of changes but introduce significant latency and context-switching for the developer.
\end{itemize}

The goal of a Smart Paste feature is to address these shortcomings by developing a fast in-IDE suggestion feature. Past work~\cite{allamanis2017smartpaste, liu2023adaptivepaste}, has shown that deep learning models can be used to predict such edits. However, delivering the Smart Paste concept as a production-ready IDE feature requires overcoming several significant data, modeling, and systems challenges. 

\begin{figure}[h]
\centering
\includegraphics[width=1.0\columnwidth]{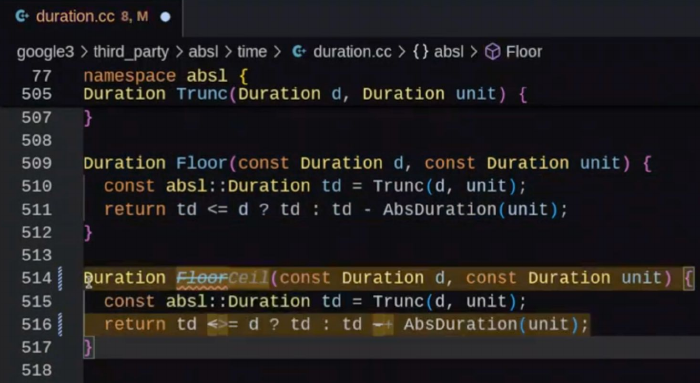}
\caption{Smart Paste monitors a developer's IDE activity, so that when a paste takes place it can render (inline) suggested edits to adapt it to the current context.}
\label{fig:smart-paste-intro}
\end{figure}

We built on this past work to develop a production-grade Smart Paste: a paste-edit suggestion feature inside Google’s IDE, now used daily by tens of thousands of developers. As shown in Figure~\ref{fig:smart-paste-intro}, the feature listens to everything pasted into the editor, then suggests fixes directly within the paste region that render as inline edits. Users can accept these suggestions with a tab or dismiss them with any other key or cursor movement.

This paper describes the practical challenges and solutions involved in building this real-world tool. Successfully productionizing this feature required addressing a range of interconnected issues across the product stack, from devising a novel data collection pipeline for the inherently ambiguous task of ``fixing'' a paste, to engineering a system that could support a diverse, multilingual code base at enterprise scale with millisecond latency, and ultimately designing an intuitive, discoverable, and non-intrusive in-editor user experience. Each of these core challenges is defined and explored in greater detail in a subsequent section. Our holistic account of this development process can guide other AI practitioners in building similar coding assistance features.

The core contributions of our work are:

\begin{itemize}
\item A detailed account of the development of a production-grade Smart Paste IDE feature, including our novel method for creating fine-tuning datasets.
\item A thorough evaluation, both offline and online, demonstrating the value of Smart Paste for Google developers at enterprise scale.
\end{itemize}

The remainder of this paper is structured as follows. We first define the problem in \S\ref{sec:problem_statement}. Next, we detail our core technical contributions, covering the data curation (\S\ref{sec:data_collection}), model training (\S\ref{sec:model-training}), and user experience (\S\ref{sec:ux}). We then evaluate our work through online and offline results (\S\ref{sec:results}) and a discussion of usage patterns (\S\ref{sec:discussion}). We conclude by examining threats to validity (\S\ref{sec:threats}), related work (\S\ref{sec:related}), and directions for future work (\S\ref{sec:future}).

%% file: problem.tex
\section{Problem Statement}
\label{sec:problem_statement}

Let a source code file consist of a sequence of tokens, $F = (t_1, t_2, \cdots, t_n)$. Let $S = (s_1, s_2,\cdots, s_m)$ be the sequence of tokens corresponding to the snippet of code that will be pasted into F. Let $[L_{\text{start}}, L_{\text{end}}]$ be the range of lines where S will be pasted. If $L_{\text{start}} == L_{\text{end}}$, pasting consists of insertion, while if $L_{\text{start}} < L_{\text{end}}$, pasting will replace the existing contents.

Let $F^{'}$ correspond to $F$ with $S$ pasted at the target 
$[L_{\text{start}}, L_{\text{end}}]$ location. 
The task of Smart Paste consists of replacing a subset of tokens within $[L_{\text{start}}, L_{\text{end}}]$, producing $F^{''}$, such that the pasted content matches the surrounding context’s syntax and semantics.

This definition of Smart Paste is similar to that presented in
past work~\cite{allamanis2017smartpaste, liu2023adaptivepaste}, but critically, we allow for unrestricted token replacements, whereas previous approaches were often limited to variable names. Tackling this problem statement at scale gives rise to two key challenges:

\boldlead{High-Quality Suggestions.} Generating useful suggestions requires a model that can navigate the inherent ambiguity of the paste editing task while maintaining high predictive accuracy across the many programming languages used in Google's monorepo.

\boldlead{Productionization Challenges.} Beyond predictive accuracy, a successful deployment of Smart Paste at Google requires serving low-latency suggestions within a non-intrusive interface that allows developers to easily review edits. This builds the developer trust essential for achieving a high acceptance rate~\cite{brown2024identifying}.

In the remainder of this paper we discuss how we addressed these two core challenges to deliver a feature that is used daily by thousands of Google developers.

%% file: data.tex
\begin{figure*}[!ht]
\small
\centering
\includegraphics[width=1.0\textwidth]{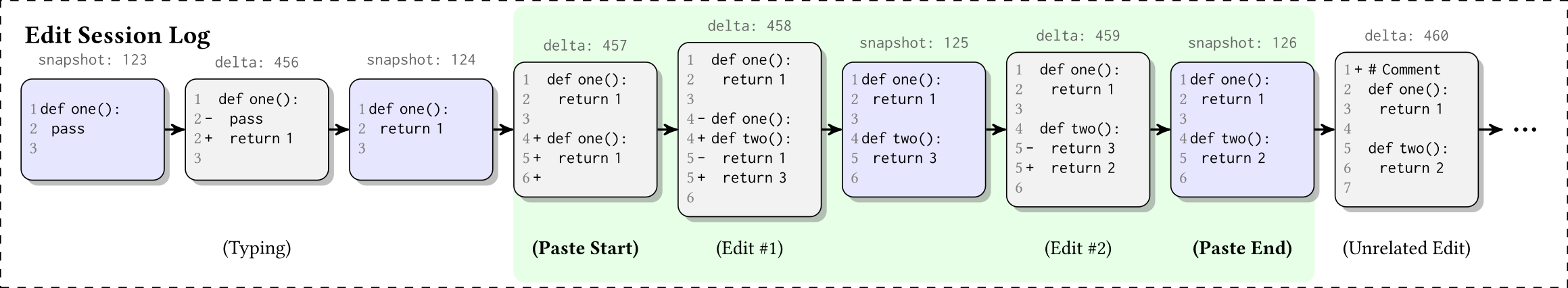}
\caption{A developer's coding journey, reconstructed from File Snapshot and Edit Delta events. Our method identifies a Paste Start and tracks related edits to find the final Paste End state. The goal of Smart Paste is to predict this final state given only the initial paste.}
\label{fig:edit-session}
\end{figure*}
\section{Data Collection}
\label{sec:data_collection}

We first focus our discussion on data collection, which was essential for addressing the core product challenges of achieving \emph{high-quality suggestions} and \emph{robust multilingual support}.

A primary hurdle was the availability of relevant data. While a small dataset was sufficient for initial feasibility tests, our early explorations demonstrated that existing models produced mixed results and were insufficient for a reliable, low-latency production feature. This analysis made it clear that achieving the required accuracy demanded a pivot to a specialized fine-tuned model (discussed further in \S\ref{sec:model-training}), which fundamentally scaled the data collection challenge from tens of examples to potentially hundreds of thousands.

This new scale required a data pipeline that could capture real-world developer copy/paste edit behaviors across many programming languages. Prior research~\cite{bavarian2022efficient} suggested we could generate this via synthetic datasets and fill-in-the-middle representations, an approach successfully used on previous features. In this approach, we would artificially create a set of ``incorrect'' or ``incomplete'' pastes and make the model predict the original correct code. However, at Google, most developers use an internal IDE called Cider, which provided us with a unique opportunity to collect real, detailed data about their coding process. 

Thus, we evolved our logging system to record the entire coding journey (illustrated in Figure~\ref{fig:edit-session}) using two main entry types:

\begin{enumerate}
\item \textbf{File Snapshot}: captures the file content at a specific point in time such as on saving or when closing the IDE.

\item \textbf{Edit Delta}: records every small change made after a snapshot.
\end{enumerate}

Even though our new logging system could reconstruct edit history, a big challenge remained: the logs did not explicitly specify an edit's provenance; for example, whether an edit came from a paste, a code completion, or manual typing. Figuring out if something was a paste was not straightforward and remained a technical hurdle. We now discuss our rule-based solution to this problem.

\subsection{Identifying Paste and Related Edit Actions}
To create our dataset, we developed a rule-based method to identify \emph{paste and fix} sequences from raw edit logs. First, we identify a \emph{paste candidate} as a bulk insert of 5-10 characters spanning 1-5 non-empty lines, distinguishing it from regular typing.

Once a paste candidate is identified, we define the subsequent \emph{fix} as any follow-up edits that start within the range of the original paste. Using this location-based heuristic, our analysis shows that 72\% of all paste events receive a local fix. Our edit tracking stops as soon as an ``unrelated edit'' occurs outside this region (see Figure \ref{fig:edit-session}). However, this heuristic can be imperfect; it may prematurely stop tracking if a user momentarily edits another part of the file before returning to finish fixing the paste (we made an exception for edits that added new imports, which often occur outside the pasted code). Consequently, while not every collected example can be guaranteed to represent the user's complete and final fix, we mitigate this risk by collecting millions of examples.

Manual checks on approximately 200 samples showed that over 80\% were plausible paste-and-fix sequences, defined as edits a human would judge as reasonable without performing compilation or execution checks. This fulfilled the main goal of our data collection heuristic: to capture data that gets the developer significantly closer to their intended state, aligning with Smart Paste's primary goal of reducing the keystrokes required to integrate pasted code.

\subsection{No-Edit Pastes}
\begin{figure}[!ht]
\centering
\includegraphics[width=0.45\textwidth]{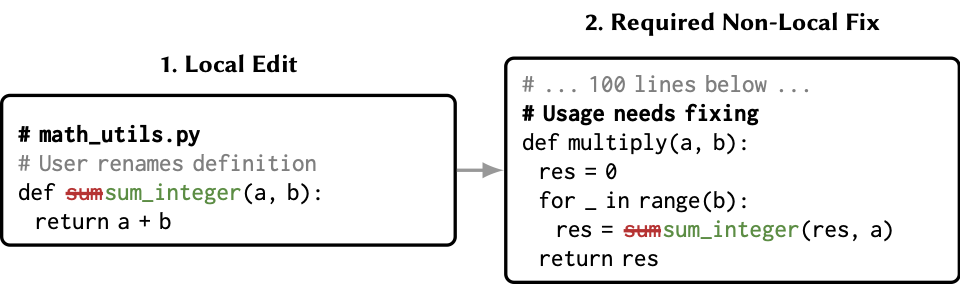}
\caption{Non-local fix scenario. (1) The user renames \texttt{sum} to \texttt{sum\_integer}. (2) This breaks the usage in \texttt{multiply}, which requires a fix outside the current view.}
\label{fig:non-local-fix-renames}
\end{figure}

To improve suggestion quality, the model must learn when to remain silent. Initial versions trained exclusively on pastes with ``fix-applied'' outputs were overly aggressive, resulting in low user acceptance. Returning to the data, we found that 28\% of paste events resulted in no local follow-up edits. This subset includes both perfect pastes and those requiring fixes outside the current IDE viewport (e.g., Figure \ref{fig:non-local-fix-renames}); since our heuristic only captures local changes, we treat both cases as \emph{no-edit} examples. Incorporating these into subsequent trainings was essential to curb over-triggering and increase the user acceptance rate.

\subsection{Curating a high-quality dataset}
Our final dataset aggregates millions of code edits, with over \emph{7 million training examples across 24 different programming languages}. We applied several filters to this data to improve its quality:

\begin{itemize}
\item Pastes were limited to a maximum of 20 lines to exclude large edits and ensure suggestions fit within the IDE viewport, a UX limitation of our initial solution.

\item Third-party licensed code (identified by metadata) and code with undetermined provenance were excluded to comply with copyright and to bias suggestions towards Google's internal style guides.

\item Logs older than four months were excluded to avoid suggestions involving deprecated code and obsolete libraries.

\item We removed examples longer than 50,000 characters to filter out large data dumps or auto-generated files not representative of daily development.
\end{itemize}

%% file: training.tex
\section{Model Training}
\label{sec:model-training}

In this section, we describe the training process for teaching a transformer-based model to perform paste edits and how we addressed important challenges.
\subsection{A Unified Model}
To be successful across Google's diverse development environment—with its many programming languages and use cases (e.g., bug fixing, feature development)—Smart Paste must provide meaningful suggestions while remaining fast and easily deployable. These requirements led to two key architectural constraints: 1) the feature must be powered by a single, unified model to avoid the complexity of maintaining separate models per language, and 2) the model must be small enough to support fast inference that meets developer expectations for in-IDE suggestions. We now discuss how our modeling strategy satisfied these constraints.

\subsection{Model Choice and Task Representation}

\begin{figure}[!ht]
\centering
\includegraphics[width=0.45\textwidth]{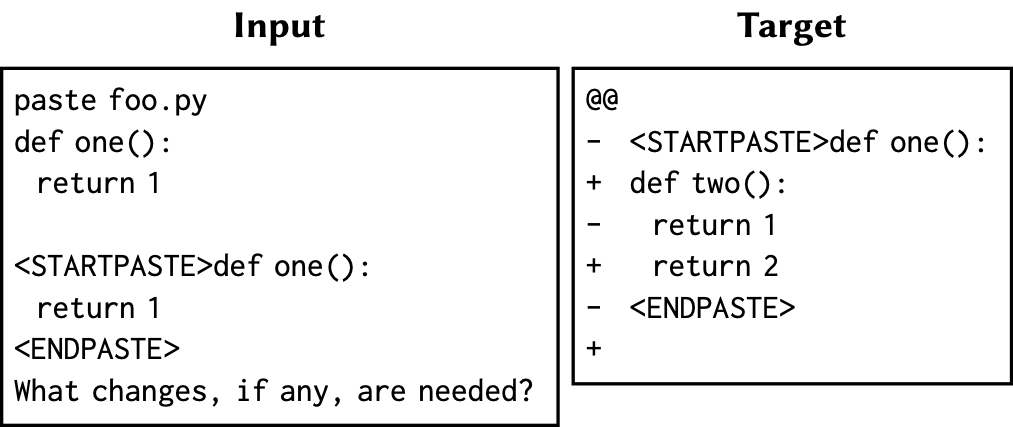}
\caption{
Task representation for the Smart Paste model using the example from Figure \ref{fig:edit-session} -- the output
unidiff format allows us to represent edit suggestions
and no-edit suggestions (not shown) in the same way.}
\label{fig:model-input-output}
\end{figure}

To meet our requirements for a single, fast, and multilingual solution, we chose to fine-tune a small version of Gemini pre-trained on Google code ~\cite{gemini2023}. To leverage this pre-training, we aligned our fine-tuning task to a familiar code transformation format. Figure \ref{fig:model-input-output} displays the input prompt, combining a task prefix ~\cite{raffel2020exploring}, code context, pasted content wrapped in special delimiter tokens, and a final instruction for the model to predict the necessary edits.

The target output is a unidiff-style~\cite{johnson1996diff} patch that uses special delimiter tokens—rather than line numbers—to apply the edit. This approach minimizes generated tokens to reduce latency and unifies the format for both edits and no-edits, which our experiments confirmed was essential for performance.

\subsection{Capturing Non-Local Context while Minimizing Latency}
While large context windows are beneficial, including entire files can introduce unacceptable latency for an in-the-flow feature like Smart Paste. We restrict input sequences to 4096 tokens, a limit chosen to balance model size against strict real-time latency thresholds. Offline testing indicated that increasing this token budget yielded no significant gains. To maximize the utility of the token budget, we strategically select relevant code local to the paste and non-local context (e.g., imports) using a greedy algorithm (Algorithm~\ref{alg:context}).

\begin{algorithm}[H]\small
    \caption{Greedy Context Selection}
    \label{alg:context}
    \begin{algorithmic}[1]
    \Statex \parbox[t]{\dimexpr\linewidth-\algorithmicindent}{
            \textbf{Input:} $L$, the sequence of file lines where $L_i$ is the $i$th line; $R$, the paste region; $B$, a token budget set to 4096 in our experiments. \\
            \textbf{Output:} $C$, a set of lines from $L$ representing the final context.
            \vspace{0.5em}
        }
        \Procedure{BuildContext}{$L, R, B$}
            \State $l_s, l_e \gets \min(R), \max(R)$ \hfill\(\leftarrow\)\text{ First and last line of paste}
            \State $C \gets \{L_0\} \cup \{L_i \mid i \in R\}$ \hfill\(\leftarrow\)\text{ Seed context}
            \If{$\text{Tokens}(C) > B$} \textbf{return} $\emptyset$ \EndIf
            
            \State $p_h, p_p, p_s \gets 1, l_s-1, l_e+1$ \hfill\(\leftarrow\)\text{ Initial expansion pointers}
            
            \Loop
                \State $C_{prev} \gets C$ \hfill\(\leftarrow\)\text{ Snapshot context to detect changes}
                \If{$L_{p_h} \notin C \land \text{Tokens}(C \cup \{L_{p_h}\}) \le B$}
                    \State $C \gets C \cup \{L_{p_h++}\}$
                \ElsIf{$L_{p_p} \notin C \land \text{Tokens}(C \cup \{L_{p_p}\}) \le B$}
                    \State $C \gets C \cup \{L_{p_p--}\}$
                \ElsIf{$L_{p_s} \notin C \land \text{Tokens}(C \cup \{L_{p_s}\}) \le B$}
                    \State $C \gets C \cup \{L_{p_s++}\}$
                \EndIf
                \If{$C = C_{prev}$} \textbf{break} \EndIf \hfill\(\leftarrow\)\text{ Exit if no lines were added this round}
            \EndLoop
            \State \textbf{return} $C$
        \EndProcedure
    \end{algorithmic}
\end{algorithm}

This process begins by seeding the context ($C$) with the paste region ($R$) and the file's first line ($L_0$). We expand context using three pointers: $p_h$ moving down from the file start (header), $p_p$ moving up from the paste start (prefix), and $p_s$ moving down from the paste end (suffix), until the budget ($B=4096$) is met or the entire file has been included. The termination condition $C = C_{prev}$ indicates that the context has stabilized, whether by exhausting the budget or capturing all available lines. The pointer starting from the top ($p_h$) is crucial for capturing headers, which contain critical information such as library imports. This context-building approach is used consistently during both training and inference.

\subsection{Multilingual Training}
We utilized a model pre-trained on Google code to support the proprietary languages and libraries of our engineering environment. Initial fine-tuning on just Python showed strong transfer to general-purpose languages—likely due to pre-training and shared paste-edit patterns—but failed on domain-specific languages like CSS. Therefore, our final approach utilized a mixed dataset spanning 24 languages, weighted by Google developer paste frequency over a one-month period. This strategy improved performance universally without regressing on Python.

\subsection{Data Batching}
During training, we observed that the configuration of training batches was important. Specifically, we included a mix of Smart Paste tasks that require edits and those that do not require edits – the latter account for approximately 30\% of examples per batch. Batches also include a mix of language examples. Both of these resulted in larger batch sizes (32 and 64) and resulted in better performance.

\subsection{Inference}
To make suggestions for new paste instances, we initially scored each potential output diff by its probability under the trained model and used a threshold to reject low-probability suggestions. However, we found that gating suggestions based on probability was no longer necessary due to model improvements and so later model versions allowed us to remove the threshold altogether.

%% file: ux.tex
\section{User Experience and Interface Design}
\label{sec:ux}
The Smart Paste user experience must be discoverable, non-intrusive, and keep developers in a ``flow state'' while clearly presenting changes. These criteria guided our design through several iterations
to produce our final user interface. We discuss alternative
designs and their tradeoffs in detail in Section~\ref{sec:rq4}.

\subsection{Inline Diff Suggestions}
We chose to render suggestions as inline code diffs, italicizing the entire suggestion, with insertions shown in gray and deletions marked by a strike-through. For highlighting, we applied a yellow tint to the entire modified line. This single ``alert'' color—unlike standard green (ok) or red (error)—draws user attention to all changes communicating that the suggestion requires careful inspection. While this approach had known limitations, such as an inability to render new-line additions (as decorations only work on existing lines) and difficulty interpreting large changes, an internal study confirmed it was still developers' preferred design, striking an effective compromise for an inline diff experience.

Interactions mirror standard IDE behaviors to minimize cognitive load: \texttt{TAB} accepts the suggestion and \texttt{ESC} dismisses it. Crucially, the suggestions do not auto-dismiss based on time; if the user performs no action (e.g., reading the code), the suggestion remains visible to allow for a thorough review.

\begin{figure}[h]
\centering
\includegraphics[width=\columnwidth]{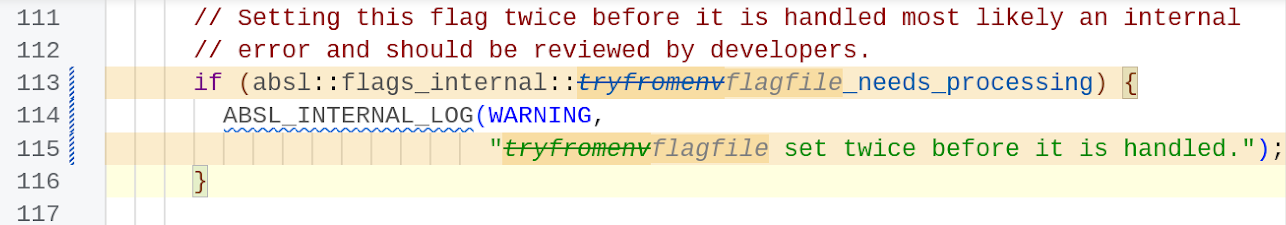}
\caption{Our user interface renders suggestions as an inline diff with yellow highlighting. This design allows for easy review, which is critical for building the developer trust needed to drive feature adoption.}
\end{figure}

Initially, to ensure a non-intrusive experience, we hid suggestions immediately if the user performed any non-typing events such as moving the cursor or scrolling out of view. However, users frequently reported accidental dismissals. We refined this behavior by delaying the dismissal from non-typing events so that a suggestion is shown for at least 2 seconds. This small change yielded a surprising 30.4\% increase in our acceptance rate, demonstrating that the original ephemeral design was preventing many users from acting on valid suggestions.

Finally, to address early tester frustration with suggestions that deleted entire pasted snippets, we implemented a post-processing rule to suppress full-deletion outputs while still permitting partial deletions.

\subsection{Low Latency Suggestions}
A key challenge for Smart Paste is delivering suggestions quickly enough to be useful without disrupting the user's workflow. We identified that the primary latency bottleneck was the model's prefill stage (processing the input), which stems from the asymmetric nature of the task: the input code context is typically much larger than the small output diff. To mitigate this, we deployed our model using a disaggregated serving infrastructure that separates the compute-intensive prefill stage from the less demanding decode stage. Specifically, we configured our serving slices with a 2:1 prefill:decode ratio, effectively dedicating twice the compute capacity to processing the long input prompt and significantly reducing end-to-end latency.

%% file: results.tex
\section{Results}
\label{sec:results}
We evaluate the Smart Paste feature at Google by addressing the following research questions:
\begin{itemize}
    \item RQ1: To what extent does the Smart Paste feature deliver effective suggestions to Google developers?
    \item RQ2: Does Smart Paste perform effectively across multiple languages in production, and what is the impact of multilingual training data?
    \item RQ3: To what extent can the Smart Paste model effectively distinguish between cases that require suggestions and no-edit pastes, which require no further changes?
    \item RQ4: What are alternative user interface designs for Smart Paste and what are their downsides?
\end{itemize}

Here, RQ1-RQ3 directly evaluate our first core challenge of delivering high quality suggestions across a multilingual monorepo. RQ1 also addresses productionization challenges by measuring the effectiveness of the end-to-end system. RQ4 provides a deep dive into the user interface design trade-offs.

To answer these questions, we rely on three primary sources of data: (1) Google's production telemetry, measuring real-world feature usage; (2) a held-out test set of 73,000 paste examples, carefully constructed to match the training set's language distribution and prevent file-path overlap; and (3) qualitative feedback on user interface experiences. A detailed definition of our evaluation metrics is presented
in Table~\ref{tab:eval-metrics}.

\begin{table}[!h]
\small
\centering
\begin{tabular}{@{} p{0.20\columnwidth}p{0.75\columnwidth}@{}}
\toprule
\textbf{Metric} & \textbf{Definition} \\
\midrule
\multicolumn{2}{c}{\textit{\textbf{Online Telemetry}}} \\
\midrule
Acceptance Rate & The fraction of suggestions accepted by users. \\
\addlinespace
Average Chars Modified & Average characters changed per accepted suggestion, measured by Longest Common Subsequence (LCS) distance. \\
\addlinespace
Average Chars Added & Average new characters added per accepted suggestion.
\\
\addlinespace
Survival & The mean ratio of added characters that stayed in the file 30 minutes after the paste. \\
\addlinespace
Throughput & The queries per second (QPS) served. \\
\addlinespace
Latency & The total time from the paste event to the suggestion being rendered in the IDE. \\
\midrule
\multicolumn{2}{c}{\textit{\textbf{Offline Test Set}}} \\
\midrule
Exact Match & The percentage of model predictions that perfectly matched the ground-truth fix.
\\
\addlinespace
chrF & Median character 6-gram F-score that measures partial overlap with the ground truth. We computed only on incorrect predictions to assess the quality of near-misses.

 \\
\addlinespace
Recall & The fraction of paste instances that require an edit where our model predicts a suggestion.
\\
\bottomrule
\end{tabular}
\caption{
To assess the performance of our system, we used a suite of metrics divided into two categories: online telemetry and offline test set evaluation.
}
\label{tab:eval-metrics}
\end{table}

\subsection{RQ1: To what extent does the Smart Paste feature deliver effective suggestions to Google developers?}
We assess the feature's effectiveness in Google’s IDE (Cider) using online deployment metrics from a stable four-month period. This timeframe was selected to avoid confounding variables from other significant infrastructure changes.

The key performance indicators shown in Table~\ref{tab:online_metrics_combined} highlight the feature's successful deployment. Developers accepted 45\% of all suggestions, and each acceptance saved them significant effort by modifying an average of 22 characters—a metric that captures the full scope of transformations including deletions and replacements, not just additions. Our survival metric demonstrates the durability of these edits: 58\% of characters added by Smart Paste remain in the file after 30 minutes, which compares favorably to the ~54\% survival rate for standard code completion. At Google's enterprise scale, this impact is substantial: \emph{accepted characters from Smart Paste account for over 1\% of all code written company-wide}.

Critically for an in-IDE tool, suggestions were delivered rapidly. The median end-to-end latency was just 346 ms, which includes 257 ms for model inference and $\sim$100 ms for system overhead. The system also proved capable of handling enterprise-scale demand, with a peak throughput of over 20 queries per second (QPS).

\begin{table}[h]
\centering
\begin{tabular}{@{} l c c c @{}}
\toprule
\textbf{Metric} & \textbf{Cider} & \textbf{SQL-IDE} & \textbf{Colab} \\
\midrule
Acceptance Rate & 45\% & 49\% & 35\% \\
\addlinespace
Avg. Chars Modified & $\sim$22 & $\sim$18 & $\sim$20 \\
\addlinespace
Avg. Chars Added & $\sim$15 & - & - \\
\addlinespace
Survival (30 min) & $\sim$58\% & - & - \\
\addlinespace
Throughput (peak) & $>$20 QPS & $>$5 QPS & $>$2 QPS \\
\addlinespace
Latency (median) & 346 ms & 350 ms & 350 ms \\
\bottomrule
\end{tabular}
\caption{Comparison of key online performance metrics for Smart Paste across three distinct developer environments: the primary Cider IDE, the SQL scripting tool, and the Colab notebook environment. These results highlight consistent value (20+ chars per acceptance) and low latency (350 ms).}
\label{tab:online_metrics_combined}
\end{table}

With these measurements, we conclude that (1) Smart Paste suggestions are actively used by Google developers, (2) these suggestions save them a substantial number of keystrokes, (3) the serving infrastructure is able to fulfill both high throughput and low latency requirements needed for an in-IDE feature.

\leadbold{Generalizing to Other Surfaces.} To validate the generalizability of our approach, we integrated Smart Paste into two distinct developer platforms: Colab (a Python notebook environment) and an internal tool for SQL-like scripting. Critically, both integrations leveraged the same underlying model and backend infrastructure, requiring only frontend work to visualize the suggestions. As shown in Table~\ref{tab:online_metrics_combined}, the feature performed strongly on both surfaces. The SQL-IDE achieved a significantly higher acceptance rate (49\%) than Colab (35\%), though the core utility was consistent, with both saving developers around 20 characters per acceptance at the same low latency. Average characters added and 30-min survival were not collected because these frontend integrations were scoped to prioritize characters modified as the core measure of user effort saved. 

We hypothesize that the discrepancy in acceptance rates between these environments is attributable to the structural complexity of the Colab notebook environment. While our method of concatenating all cells provided the model with a single, unified context, the multi-cell format introduced challenges on the frontend. Applying a suggestion that added or removed lines could require dynamically expanding or contracting a cell's boundaries, which is a more complex rendering task than editing a standard text file. This frontend complexity, in contrast to the self-contained and structured nature of SQL scripts, likely contributed to a lower acceptance rate in Colab. Despite the unique frontend challenges presented by the notebook environment, this successful integration into diverse IDEs still demonstrates a key architectural advantage of our design: a single, powerful model backend can provide a consistent AI feature across a diverse ecosystem of tools, maximizing impact with minimal incremental engineering cost.

\subsection{RQ2: Does Smart Paste perform effectively across multiple languages in production, and what is the impact of multilingual training data?}
For Smart Paste to be a successful feature, it must deliver high-quality suggestions across the many programming languages used at Google. As shown in Table 
\ref{tab:online_metrics_survival}, online production metrics confirm that the feature achieves consistently high acceptance and 30-minute survival rates across a wide variety of languages.

We find that the feature achieves high acceptance rates for core languages like C++ (47.8\%), Java (44.6\%), and Python (45.4\%). This strong performance extends across the board, as the diverse, aggregated groups of less common languages also show consistently high acceptance rates, proving the feature's tangible value to developers working in different parts of the codebase.

\begin{table}[ht!]
\centering
\setlength{\tabcolsep}{5pt} 

\label{tab:online_metrics_no_percent}
\begin{tabular}{lcc}
\toprule
 & \textbf{Acceptance (\%)} & \textbf{Survival (\%)} \\
\midrule
\textbf{Popular Languages} \\
\cmidrule(r){1-1} 
C++                          & 47.8 & 59.1 \\
Java                         & 44.6 & 58.3 \\
Python                       & 45.4 & 60.3 \\
\midrule
\textbf{Less Common Languages} \\
\cmidrule(r){1-1} 
Build \& Config.       & 42.7 & 55.4 \\
Data \& Query              & 40.5 & 65.8 \\
General-Purpose              & 44.4 & 55.9 \\
Web Development              & 42.5 & 56.9 \\
\bottomrule
\end{tabular}
\caption{
Online acceptance rate and survival metric for Smart Paste in Cider
over 16 high-volume languages (groups are aggregated with average) show that a high fraction of suggestions are accepted and a large portion of the resulting characters survive for over 30 minutes.
}
\label{tab:online_metrics_survival}
\end{table}

Recall from Section~\ref{sec:model-training}, that our Smart Paste model was trained on multilingual data. To measure the impact of multilingual training data, we carried out an offline evaluation where we compare the base Smart Paste model with one trained exclusively on Python. This analysis was performed offline, as our goal is always to deploy the best-performing model to users.

As summarized in Table \ref{tab:offline_metrics_comparison},  a model fine-tuned only on Python achieves a respectable cross-language capability, with a 45.1\% overall exact match from our initial data. However, training on multilingual data provides a significant and consistent performance lift across all language categories. We do not observe any regression in Python, in fact this provides a marginal 0.8 percentage point improvement in overall exact match accuracy. The multilingual training impact is most pronounced in diverse language categories, with both the Build \& Configuration and General-Purpose groups seeing an average performance increase of 3.9 percentage points. This confirms that multilingual training is a critical factor in the feature's ability to serve a diverse developer base effectively. 

\begin{table}[ht!]
\small 
\centering
\setlength{\tabcolsep}{3pt} 

\label{tab:final_bold_headers}
\begin{tabular}{lcccccc}
\toprule
 & \multicolumn{2}{c}{\textbf{Edit}} & \multicolumn{2}{c}{\textbf{No-Edit}} & \multicolumn{2}{c}{\textbf{Overall}} \\
\cmidrule(lr){2-3} \cmidrule(lr){4-5} \cmidrule(lr){6-7}
 & Python & All & Python & All & Python & All \\
\midrule
\textbf{Popular Languages} \\
\cmidrule(r){1-1} 
C++                          & 27.9 & 31.6 & 74.9 & 78.8 & 41.8 & 45.6 \\
Java                         & 29.5 & 35.3 & 74.6 & 79.1 & 43.1 & 48.5 \\
Python                       & 29.0 & 29.7 & 81.1 & 82.3 & 47.0 & 47.8 \\
\midrule
\textbf{Less Common Languages} \\
\cmidrule(r){1-1} 
Build \& Config.       & 29.2 & 32.5 & 81.8 & 86.8 & 49.4 & 53.3 \\
Data \& Query              & 30.5 & 32.0 & 80.0 & 83.1 & 49.1 & 51.2 \\
General-Purpose              & 30.2 & 33.9 & 78.3 & 82.9 & 45.9 & 49.8 \\
Web Development              & 29.4 & 34.3 & 81.7 & 82.5 & 46.1 & 49.8 
\\
\midrule
\textbf{Overall} & 29.4 & 33.5 & 78.2 & 81.9 & 45.1 & 49.0 
\\
\bottomrule
\end{tabular}
\caption{
Exact match accuracy comparison between a Python-only model and the multilingual production model on the held-out test set shows
training on multilingual data
improves predictions.
}
\label{tab:offline_metrics_comparison}
\end{table}

\subsection{RQ3: To what extent can the Smart Paste model effectively distinguish between cases that require suggestions and no-edit pastes, which require no further changes?}

Table \ref{tab:offline_full_metrics} also shows the exact match accuracy between paste instances that require an edit and those that do not. We observe that the exact match accuracy for no-edit (i.e. cases where the model should predict an empty diff), is high across languages and groups, when trained on multilingual data. 

\begin{table}[ht!]
\small
\centering
\setlength{\tabcolsep}{4pt}

\begin{tabular}{lccccc}
\toprule
 & \multicolumn{3}{c}{\textbf{Accuracy}} & \multirow{2}{*}{\textbf{Recall}} & \multirow{2}{*}{\textbf{chrF}} \\
\cmidrule(lr){2-4}
 & Edit & No-Edit & Overall & & \\
\midrule
\textbf{Popular Languages} \\
\cmidrule(r){1-1}
C++ & 31.6 & 78.8 & 45.6 & 61.4 & 95.1 \\
Java & 35.3 & 79.1 & 48.5 & 69.0 & 95.1 \\
Python & 29.7 & 82.3 & 47.8 & 66.1 & 94.8 \\
\midrule
\textbf{Less Common Lang.} \\
\cmidrule(r){1-1}
Build \& Config. & 32.5 & 86.8 & 53.3 & 79.0 & 95.0 \\
Data \& Query & 32.0 & 83.1 & 51.2 & 74.2 & 94.7 \\
General-Purpose & 33.9 & 82.9 & 49.8 & 70.3 & 95.4 \\
Web Development & 34.3 & 82.5 & 49.8 & 69.0 & 94.8 \\
\midrule
\textbf{Overall} & 33.5 & 81.9 & 49.0 & 69.8 & 94.9 \\
\bottomrule
\end{tabular}
\caption{
Offline metrics for the multilingual model. The model recovers a majority of required edits (69.8\% recall), and the high median chrF on incorrect suggestions, indicates that even the model's mistakes have a substantial overlap with the ground-truth solution.
}
\label{tab:offline_full_metrics}
\end{table}

In addition, Table \ref{tab:offline_full_metrics} shows that the Smart Paste model is able to recover most of the paste instances that require edits – with an overall recall of 69.8\%. We also find that when the model output is incorrect based on exact match, the median chrF is high \cite{popovic-2015-chrf}. This indicates that incorrect suggestions still maintain a high degree of similarity to the ground truth, rather than generating arbitrary or irrelevant code.

Build and configuration languages demonstrate the highest overall accuracy (53.3\%), no-edit accuracy (86.8\%), and recall (79.0\%).  We attribute this to the structured syntax and common usage patterns in these files, such as string manipulations for correcting file paths.

\subsection{RQ4: What are alternative user interface designs for Smart Paste and what are their downsides?}
\label{sec:rq4}

During development, we explored several alternative UI designs to find the best way to render and interact with suggestions. We ultimately discarded each due to key trade-offs that failed to meet our core criteria of being discoverable, clear, and non-intrusive.

\begin{figure}[h]
\includegraphics[width=1.0\columnwidth]{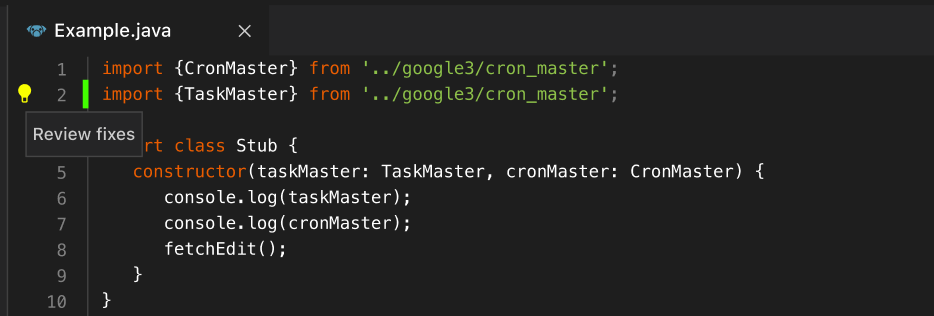}
\caption{The ``auto-apply with hint'' design automatically applied suggestions, but
this intrusive nature eroded user trust and disrupted developer flow, especially when incorrect.}
\label{fig:ux-auto-apply}
\end{figure}

\leadbold{Auto-Apply with a Hint: Convenient but Intrusive.}
We first considered an ``auto-apply'' design (Figure~\ref{fig:ux-auto-apply}), similar to auto-imports, hoping convenience would outweigh the disruption of occasional reverts. However, a user experience (UX) study found that even infrequent incorrect suggestions were highly disruptive and eroded developer trust. Participants universally preferred having control over the suggestion, a pattern consistent with other AI features in Cider. This led us to pivot to user-action-based designs.

\begin{figure}[h]
\centering
\includegraphics[width=1.0\columnwidth]{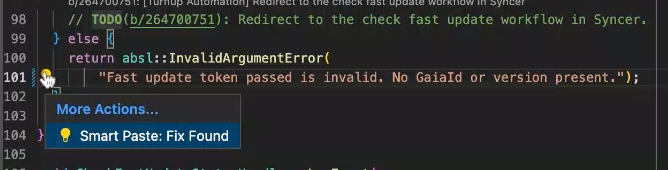}
\caption{Using the native ``light bulb'' as an entry point was non-intrusive, but suffered from poor discoverability and led to a disruptive, multi-step review process.}
\label{fig:ux-light-bulb}
\end{figure}

\begin{figure}[h]
\centering
\includegraphics[width=1.0\columnwidth]{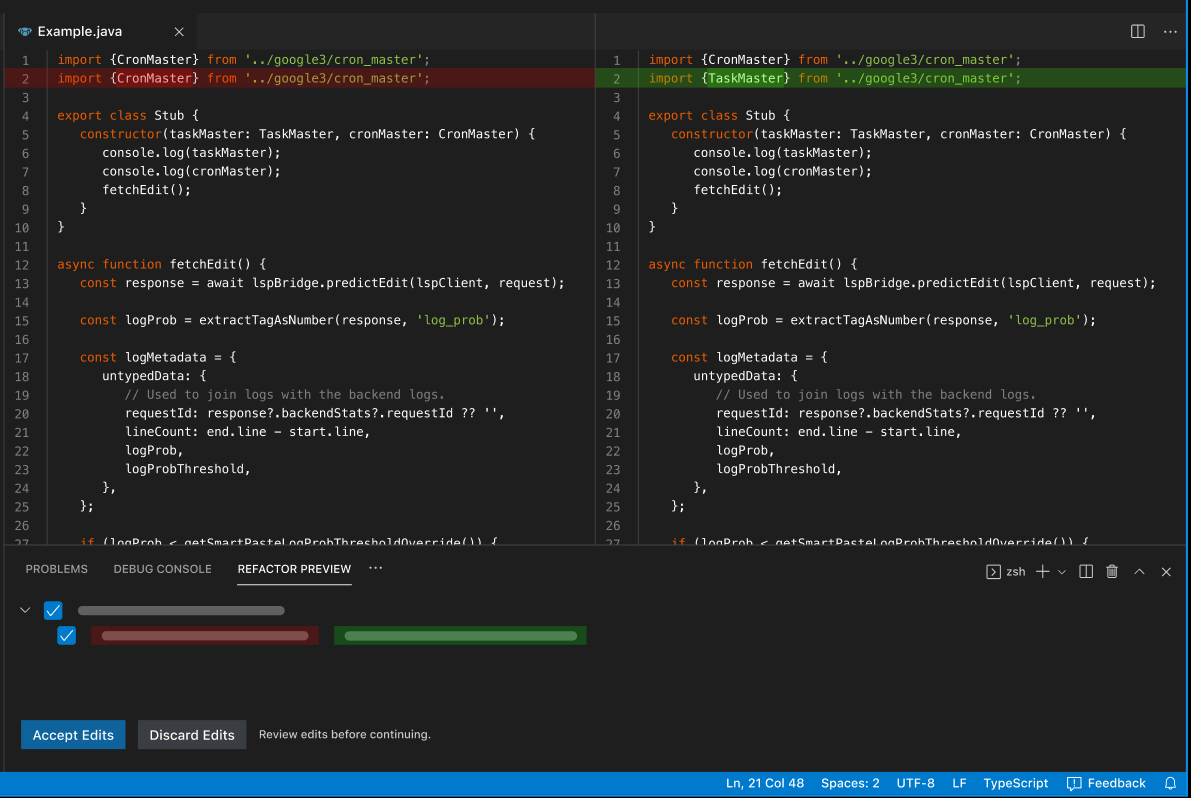}
\caption{The side-by-side diff view that resulted from the Code Action forced a context switch to a new tab for review, proving too disruptive to the developer's workflow, which was a significant barrier to adoption.}
\label{fig:ux-code-action-side-by-side}
\end{figure}

\leadbold{Code Actions: Unintrusive but Undiscoverable.}
Leveraging the native VS Code Code Actions~\cite{vscode:refactoring} (the ``light bulb'' icon) provided an unintrusive entry point (Figure~\ref{fig:ux-light-bulb}). After interaction with the light bulb, a diff view would open in a new tab, showing the change to the pasted content as seen in Figure~\ref{fig:ux-code-action-side-by-side}.

While non-intrusive, this approach was inconvenient, requiring multiple interactions (open dropdown, select ``Paste Fix'', review, accept) that disrupted the user's flow. Furthermore, UX studies for other AI features confirmed the lightbulb icon lacks discoverability.

\begin{figure}[h]
\centering
\includegraphics[width=1.0\columnwidth]{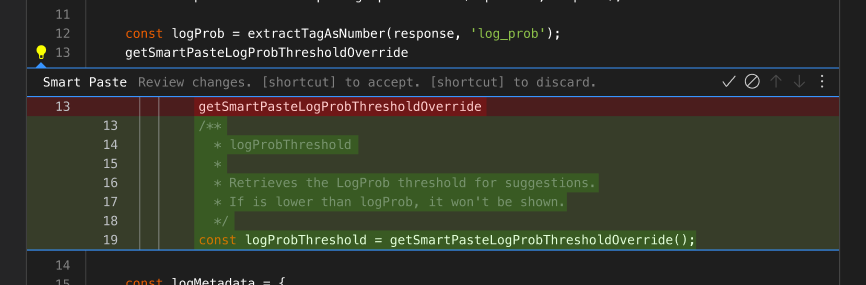}
\caption{The Peek View suggestion window. Although capable of showing complex diffs and supporting user modifications, this UI was too heavyweight and disruptive for the common case of simple paste edits, creating a poor user experience.}
\label{fig:ux-peek-suggestion}
\end{figure}

\leadbold{Peek Suggestions: Easy To Review but Disruptive.}
We also considered using VSCode Peek View~\cite{vscode:peek} to show a line-by-line diff in a window beneath the cursor, displaying insertions and deletions and allowing user modification, as shown in Figure~\ref{fig:ux-peek-suggestion}.

We ultimately discarded this approach. The flexibility for complex diffs was deemed too cumbersome for simple insertions, and we identified that most paste fixes involved small partial changes, not complete rewrites which would have benefited by this type of rendering.

\begin{figure}[h]
\centering
\includegraphics[width=1.0\columnwidth]{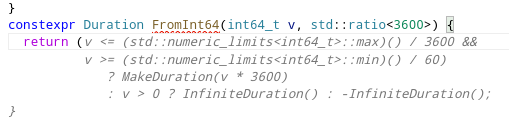}
\caption{The inline ``ghost text'' pattern, adopted from code completion was ultimately unsuitable because it could not represent replace or delete operations, fundamentally limiting the quality and scope of possible suggestions.}
\label{fig:ux-inline-ghost}
\end{figure}

\leadbold{Inline Ghosting: Lacked Replace Support.}
Another option was to leverage the pattern from the widely known and successful AI Code Completion feature~\cite{google:completion}, an in-flow experience where suggestions are rendered as grey, slanted text accepted via 
\lstinline{[TAB]} (Figure~\ref{fig:ux-inline-ghost}). 

Unfortunately, this pattern did not effectively support replace or delete operations, only inserts. This would have significantly hindered Smart Paste's functionality and conflicted with the existing code completion feature. Ultimately, we opted to adapt this pattern with major adjustments to overcome the challenge of representing replacements.

%% file: discussion.tex
\section{Practical Uses of Smart Paste}
\label{sec:discussion}

Deploying Smart Paste at an enterprise scale revealed several emergent developer behaviors and sophisticated usage patterns that evolved as users became more familiar with the feature.

\begin{figure}[h]
\centering
\includegraphics[width=1.0\columnwidth]{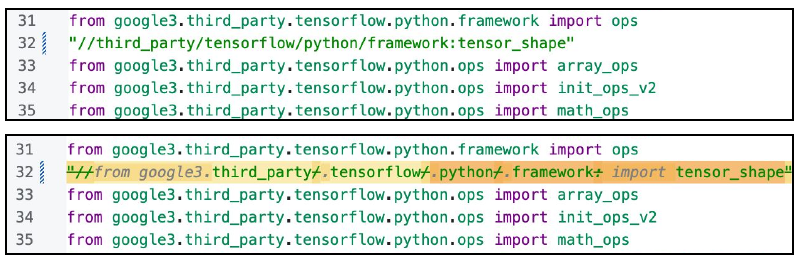}
\caption{
After deployment, users employed Smart Paste in unexpected ways, such as converting package dependencies into language imports
}
\label{fig:smart-paste-for-dependency}
\end{figure}

\leadbold{Dependency and Filepath Management.} A primary use case involved developers pasting raw file paths into code, relying on Smart Paste to automatically format them into correct import statements as shown in Figure~\ref{fig:smart-paste-for-dependency}. We also observed the reverse, where import statements were pasted into BUILD files and correctly translated into the required dependency syntax. The model proved highly effective at resolving incomplete paths, streamlining a typically tedious and manual process.

\begin{figure}[h]
\centering
\includegraphics[width=1.0\columnwidth]{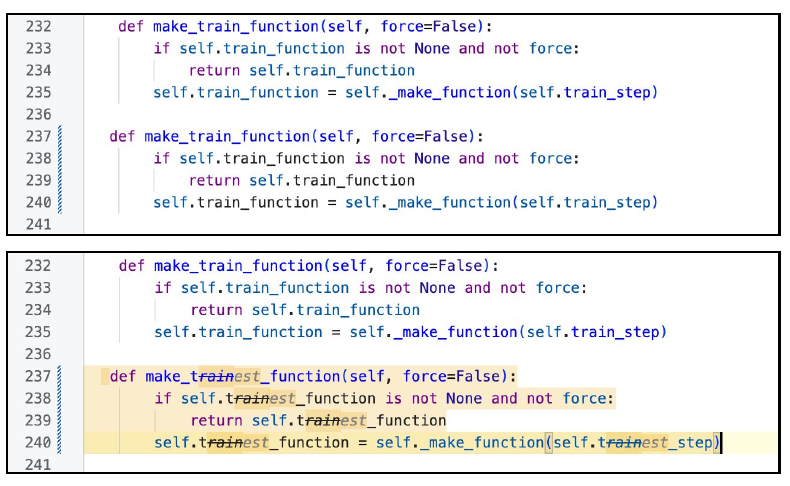}
\caption{
Developers employed chains of pastes and Smart Paste suggestions to function as a lightweight refactoring tool.
}
\label{fig:smart-paste-as-refactoring}
\end{figure}

\leadbold{Lightweight Repetitive Refactoring.} Beyond simple variable renaming—a common task cited in prior work—we observed a novel ``chaining'' behavior. Developers would execute a rapid sequence of copy-paste-accept actions to apply incremental changes across multiple lines, such as renaming sequential protobuf fields. This effectively transformed Smart Paste into a tool for lightweight, repetitive refactoring. Figure~\ref{fig:smart-paste-as-refactoring} shows one such example, where a user creates a variant of an existing method using Smart Paste.

\leadbold{API Lookup Mechanism and Quick Syntax Fixes.} Similar to code completion, developers sometimes use Smart Paste to look up common APIs without needing to open documentation or browse existing code. Specifically, developers paste incomplete or pseudo-code like fragments and rely on Smart Paste edits to make them functional. This is particularly useful across 
domain-specific code editors (e.g., Colab and SQL-IDE), where it is common to copy/paste
entire contents.
This allows developers to express their intentions in a pseudo-code format and then quickly convert the entire block into working code with just a few keystrokes.

\leadbold{Context-Aware Cross-Language Translation.} The unified multilingual model not only translated code between different languages (e.g., from one scripting language to another) but also demonstrated the crucial ability to discern when not to translate. In files containing mixed languages by design, such as Markdown with embedded Python snippets, the model correctly identified the context and preserved the pasted code's original language, avoiding erroneous transformations.

\leadbold{Workflow Adaptation and Partial Acceptance.} A key finding was that the feature's acceptance rate increased over time with no changes to the underlying model. We observed that developers began to accept suggestions that were imperfect but directionally correct, as making a small manual tweak to the suggestion was faster than fixing the original paste. We hypothesize that this ``partial acceptance'' phenomenon is evidence of users building trust, where they integrate the tool into their workflow to minimize effort rather than to achieve a perfect, one-shot fix.

%% file: threats.tex
\section{Threats to Validity}
\label{sec:threats}

The results presented in this work reflect conditions of a large software company and may not generalize to other settings (e.g. outside of enterprise environments). Similarly, many of our design choices were driven by specific requirements, such as delivering low latency paste edit suggestions. We now describe the main factors to consider for the generalizability of this work to different settings.

Collecting the dataset needed to effectively train the Smart Paste model was facilitated by keystroke-level telemetry enabled in our in-house IDE (Cider). Other enterprise settings may not have access to such granular data, and so curating a similarly high-quality dataset may require adapting or extending the heuristics presented in this work. This same telemetry allowed us to measure the impact of our feature on developers’ typing, and we used this information to evolve it. Without access to such measurements, developing a similar tool may require additional metrics for success.

We chose to train our model on paste actions in a variety of programming languages. This choice was driven by the need to serve the same feature in a large multilingual codebase. Other settings, where a single programming language dominates, may not require such multilingual training and could be simplified.

Our user interface built on prior experience with related coding assistance features, such as code completion. Importantly, users could receive Smart Paste suggestions through a familiar interface and did not need to be aware of completely new modes of interaction. Releasing a Smart Paste feature in a different editor may require a different user interface to accommodate existing code features. Indeed, as we discuss in our evaluation section, we generalized Smart Paste to a computational notebook and a SQL-like IDE by reimplementing the frontend components, applying lessons learned from our main experience.

%% file: relatedwork.tex
\section{Related Work}
\label{sec:related}

Code completion is one of the most popular and successful applications of large language models for code authoring. Industrial code completion offerings include Github Copilot \cite{github2022copilot}, Amazon CodeWhisperer \cite{aws2023codewhisperer}, and Google Gemini Code Assist \cite{google2023duetai}, among others. While classical code completion has typically been based on a combination of formal program analysis~\cite{perelman2012type} and statistical modeling~\cite{raychev2014code}, most modern code completion engines now employ LLM-based predictions and have shown to scale to whole-line and multi-line code suggestions. Such LLM-based code completion features have been successfully employed in industrial settings~\cite{dunay2024multi}. Similar to this line of work, we employ an LLM to provide real-time code authoring assistance. We share challenges such as the need for low-latency responses. However, in contrast to code completion, which typically suggests only a suffix of the current code, our work is focused specifically on automatically integrating copy/paste code into a new context, which may require edits beyond suffix completion. We showed that this interaction raises unique user experience challenges that led us to exploring diverse interface designs.

LLMs have previously been successfully used to perform targeted code edits. Tufano et al.~\cite{tufano2019learning} framed code editing as a neural machine translation task. 
Chakraborty and
Baishakhi~\cite{chakraborty2021multi} treat code editing as a machine translation task, but provide a natural language hint as an additional modality. Nam et al. study this type of natural language-based code transformation within Google~\cite{nam2025prompting}, based on a popular internal code transformation feature. Similarly, Smart Paste is an LLM-based code editing feature. However, Smart Paste is scoped to paste events, only edits paste regions, and does not use any additional hints, instead relying exclusively on the model's training and the context of the paste to suggest edits.

Closely related to adaptive source code pasting, Amidon et al.~\cite{amidon2015program} introduced the concept of code transfer within the context of program repair. Their work shows how to automatically replace or combine code from multiple applications to eliminate program defects. Sidiroglou et al.~\cite{sidiroglou2017codecarboncopy} employed a combination of dynamic and static analysis to further refine the code transplanted between two programs. Our work similarly explores how to adapt code to a new context, adjusting expressions as necessary. In contrast to code transfer work, our use case is real-time code authoring assistance, which limits the use of techniques from dynamic analysis, and we instead rely on the predictive power of LLMs.

Allamanis and Brockschmidt~\cite{allamanis2017smartpaste} defined Smart Paste as a standalone code assistance task and showed that deep learning could be used to successfully adapt variable usage in C\# code snippets pasted into a new context program. Liu et al.~\cite{liu2023adaptivepaste} integrated a similar form of intelligent copy-paste -- using a transformer model pre-trained on a dataflow-based task -- into an IDE.
Their evaluation showed this approach could adapt Python code snippets with high accuracy and a user study found that this feature could reduce the time spent performing code copy/paste. 

Our work builds on this line of research and shows that Smart Paste can be successfully delivered as an actively used feature in an industrial context. Doing so led to addressing multiple key challenges not previously tackled. First, our Smart Paste work had to support authoring in many programming languages, while past work had focused their evaluation C\# and Python. Second, a smooth user experience is critical for deployed features used by tens of thousands engineers, and so our design had to address production serving constraints and consider user interface concerns. Third, we report on a wide range of deployment-based statistics, showing the uptake of Smart Paste among professional developers.

% New content based on review feedback
We believe the ideas and design principles behind Smart Paste
can be broadly applicable, including
to code editors in the public domain. However, our specific
implementation and experiments
(including training) are uniquely tied to Google's infrastructure.

%% file: futurework.tex
\section{Future Work}
\label{sec:future}

We envision evolving Smart Paste into a comprehensive, context-aware assistant through a series of milestones.

\leadbold{Milestone 1: Expanding Suggestion Scope.}
We aim to expand suggestions beyond the paste boundary, generating follow-up edits in surrounding code, dependent files, or unit tests after the initial paste fix is accepted. This expands the feature from a localized ``paste fixer'' to an integration assistant.

\leadbold{Milestone 2: Next Edits for Broader Events.}
We plan to generalize triggers to other high-intent events, such as code completions or refactoring, offering suggestions when the developer is already evaluating their next step in a manner that minimizes cognitive disruption and maintains their workflow momentum.

\leadbold{Milestone 3: Proactive, Model-Initiated Suggestions.}
We aim to evolve the system beyond being purely reactive to specific user actions. Instead, the model will learn to independently recognize opportune moments within the workflow, anticipating user needs based on the evolving code context.

\leadbold{Milestone 4: The Next-Action Assistant.}
We plan to ultimately support holistic ``next actions'' that encompass the entire development process. This extends beyond code edits to include executing tools like builds or tests or triggering autonomous agents for multi-step tasks. This aligns our feature with recent advancements in agentic coding found in modern IDEs like Cursor \cite{cursor} and Windsurf \cite{windsurf}, where the user has less control but more complex tasks can be accomplished.

%% file: conclusion.tex
\section{Conclusion}
Copy-pasting is fundamental to software development, yet developers must often adapt snippets to their surrounding context. In this work, we showed how we trained and deployed a Smart Paste feature within Google. We share how we overcame key challenges to our deployment, including details on our multilingual data collection, model training and serving choices, and user interaction design. Ultimately, this resulted in the current deployment of this popular AI-based IDE feature at Google, with 45\% of suggestions accepted by developers, saving them on average 22 keystrokes.

%% file: acknowledgement.tex
% \section{Acknowledgements}
\begin{acks}
We thank our colleagues at Google, particularly Miltos Allamanis, Boris Bokowski, Satish Chandra, Cristopher Claeys, Madhura Dudhgaonkar, Alberto Elizondo, Simone Forte, Michael Golahi, Sandeep Katragadda, Ugam Kumar, Pascal Lamblin, Katie Le, Damien Martin-Guillerez, Henrik Muehe, Ambar Murillo, Stoyan Nikolov, Aditya Pandey, Siddhant Sanyam, Christian Schneider, Pavel Sychev, Niranjan Tulpule, Ilya Tzoop, Andy Yankovsky, and David Zhang.
\end{acks}

%% file: references.bib
@inproceedings{perelman2012type,
  title={Type-directed completion of partial expressions},
  author={Perelman, Daniel and Gulwani, Sumit and Ball, Thomas and Grossman, Dan},
  booktitle={Proceedings of the 33rd ACM SIGPLAN conference on Programming Language Design and Implementation},
  pages={275--286},
  year={2012}
}

@inproceedings{raychev2014code,
  title={Code completion with statistical language models},
  author={Raychev, Veselin and Vechev, Martin and Yahav, Eran},
  booktitle={Proceedings of the 35th ACM SIGPLAN conference on programming language design and implementation},
  pages={419--428},
  year={2014}
}

@inproceedings{dunay2024multi,
  title={Multi-line ai-assisted code authoring},
  author={Dunay, Omer and Cheng, Daniel and Tait, Adam and Thakkar, Parth and Rigby, Peter C and Chiu, Andy and Ahmad, Imad and Ganesan, Arun and Maddila, Chandra and Murali, Vijayaraghavan and others},
  booktitle={Companion Proceedings of the 32nd ACM International Conference on the Foundations of Software Engineering},
  pages={150--160},
  year={2024}
}

@inproceedings{amidon2015program,
  title={Program fracture and recombination for efficient automatic code reuse},
  author={Amidon, Peter and Davis, Eli and Sidiroglou-Douskos, Stelios and Rinard, Martin},
  booktitle={2015 IEEE High Performance Extreme Computing Conference (HPEC)},
  pages={1--6},
  year={2015},
  organization={IEEE}
}

@inproceedings{sidiroglou2017codecarboncopy,
  title={CodeCarbonCopy},
  author={Sidiroglou-Douskos, Stelios and Lahtinen, Eric and Eden, Anthony and Long, Fan and Rinard, Martin},
  booktitle={Proceedings of the 2017 11th Joint Meeting on Foundations of Software Engineering},
  pages={95--105},
  year={2017}
}

@article{allamanis2017smartpaste,
  title={Smartpaste: Learning to adapt source code},
  author={Allamanis, Miltiadis and Brockschmidt, Marc},
  journal={arXiv preprint arXiv:1705.07867},
  year={2017}
}

@inproceedings{liu2023adaptivepaste,
  title={Adaptivepaste: Intelligent copy-paste in IDE},
  author={Liu, Xiaoyu and Jang, Jinu and Sundaresan, Neel and Allamanis, Miltiadis and Svyatkovskiy, Alexey},
  booktitle={Proceedings of the 31st ACM Joint European Software Engineering Conference and Symposium on the Foundations of Software Engineering},
  pages={1844--1854},
  year={2023}
}

@article{bavarian2022efficient,
  title={Efficient training of language models to fill in the middle},
  author={Bavarian, Mohammad and Jun, Heewoo and Tezak, Nikolas and Schulman, John and McLeavey, Christine and Tworek, Jerry and Chen, Mark},
  journal={arXiv preprint arXiv:2207.14255},
  year={2022}
}

@inproceedings{ray2013detecting,
  title={Detecting and characterizing semantic inconsistencies in ported code},
  author={Ray, Baishakhi and Kim, Miryung and Person, Suzette and Rungta, Neha},
  booktitle={2013 28th IEEE/ACM International Conference on Automated Software Engineering (ASE)},
  pages={367--377},
  year={2013},
  organization={IEEE}
}

@misc{vscode:refactoring,
  title        = {Refactoring - Visual Studio Code},
  author       = {Microsoft},
  year         = {2025},
  month        = {Sep},
  howpublished = {\url{https://code.visualstudio.com/docs/editing/refactoring}},
  note         = {Accessed: 2025-09-18}
}

@misc{vscode:peek,
  title        = {Navigate and Edit C\# - Visual Studio Code},
  author       = {Microsoft},
  url          = {https://code.visualstudio.com/docs/csharp/navigate-edit#_peek-definition},
  note         = {Section: Peek Definition. Accessed: 2025-09-18}
}

@misc{google:completion,
  title        = {ML-Enhanced Code Completion Improves Developer Productivity},
  author       = {Maxim Tabachnyk and Stoyan Nikolov},
  year         = {2022},
  month        = {Jul},
  howpublished = {\url{https://research.google/blog/ml-enhanced-code-completion-improves-developer-productivity/}},
  note         = {Accessed: 2025-09-18}
}

@inproceedings{bruch2009learning,
  title={Learning from examples to improve code completion systems},
  author={Bruch, Marcel and Monperrus, Martin and Mezini, Mira},
  booktitle={Proceedings of the 7th joint meeting of the European softw
             are engineering conference and the ACM SIGSOFT symposium on the foundations of software engineering},
  pages={213--222},
  year={2009}
}

@inproceedings{svyatkovskiy2019pythia,
  title={Pythia: Ai-assisted code completion system},
  author={Svyatkovskiy, Alexey and Zhao, Ying and Fu, Shengyu and Sundaresan, Neel},
  booktitle={Proceedings of the 25th ACM SIGKDD international conference on knowledge discovery \& data mining},
  pages={2727--2735},
  year={2019}
}

@article{team2023gemini,
  title={Gemini: a family of highly capable multimodal models},
  author={Team, Gemini and Anil, Rohan and Borgeaud, Sebastian and Alayrac, Jean-Baptiste and Yu, Jiahui and Soricut, Radu and Schalkwyk, Johan and Dai, Andrew M and Hauth, Anja and Millican, Katie and others},
  journal={arXiv preprint arXiv:2312.11805},
  year={2023}
}

@article{achiam2023gpt,
  title={Gpt-4 technical report},
  author={Achiam, Josh and Adler, Steven and Agarwal, Sandhini and Ahmad, Lama and Akkaya, Ilge and Aleman, Florencia Leoni and Almeida, Diogo and Altenschmidt, Janko and Altman, Sam and Anadkat, Shyamal and others},
  journal={arXiv preprint arXiv:2303.08774},
  year={2023}
}

@article{yang2024swe,
  title={Swe-agent: Agent-computer interfaces enable automated software engineering},
  author={Yang, John and Jimenez, Carlos E and Wettig, Alexander and Lieret, Kilian and Yao, Shunyu and Narasimhan, Karthik and Press, Ofir},
  journal={Advances in Neural Information Processing Systems},
  volume={37},
  pages={50528--50652},
  year={2024}
}

@inproceedings{tomasdottir2017and,
  title={Why and how JavaScript developers use linters},
  author={T{\'o}masd{\'o}ttir, Krist{\'\i}n Fj{\'o}la and Aniche, Mauricio and Van Deursen, Arie},
  booktitle={2017 32nd IEEE/ACM International Conference on Automated Software Engineering (ASE)},
  pages={578--589},
  year={2017},
  organization={IEEE}
}

@inproceedings{brown2024identifying,
  title={Identifying the factors that influence trust in AI code completion},
  author={Brown, Adam and D'Angelo, Sarah and Murillo, Ambar and Jaspan, Ciera and Green, Collin},
  booktitle={Proceedings of the 1st ACM International Conference on AI-Powered Software},
  pages={1--9},
  year={2024}
}

@article{raffel2020exploring,
  title={Exploring the limits of transfer learning with a unified text-to-text transformer},
  author={Raffel, Colin and Shazeer, Noam and Roberts, Adam and Lee, Katherine and Narang, Sharan and Matena, Michael and Zhou, Yanqi and Li, Wei and Liu, Peter J},
  journal={Journal of machine learning research},
  volume={21},
  number={140},
  pages={1--67},
  year={2020}
}

@article{johnson1996diff,
  title={Diff, patch, and friends},
  author={Johnson, Michael K},
  journal={Linux Journal},
  volume={1996},
  number={28es},
  pages={2--es},
  year={1996},
  publisher={Belltown Media Houston, TX}
}

@inproceedings{chakraborty2021multi,
  title={On multi-modal learning of editing source code},
  author={Chakraborty, Saikat and Ray, Baishakhi},
  booktitle={2021 36th IEEE/ACM International Conference on Automated Software Engineering (ASE)},
  pages={443--455},
  year={2021},
  organization={IEEE}
}

@inproceedings{tufano2019learning,
  title={On learning meaningful code changes via neural machine translation},
  author={Tufano, Michele and Pantiuchina, Jevgenija and Watson, Cody and Bavota, Gabriele and Poshyvanyk, Denys},
  booktitle={2019 IEEE/ACM 41st International Conference on Software Engineering (ICSE)},
  pages={25--36},
  year={2019},
  organization={IEEE}
}

@article{nam2025prompting,
  title={Prompting llms for code editing: Struggles and remedies},
  author={Nam, Daye and Omran, Ahmed and Murillo, Ambar and Thakur, Saksham and Araujo, Abner and Blistein, Marcel and Fr{\"o}mmgen, Alexander and Hellendoorn, Vincent and Chandra, Satish},
  journal={arXiv preprint arXiv:2504.20196},
  year={2025}
}

@inproceedings{popovic-2015-chrf,
    title = "chr{F}: character n-gram {F}-score for automatic {MT} evaluation",
    author = "Popovi{\'c}, Maja",
    editor = "Bojar, Ond{\v{r}}ej  and
      Chatterjee, Rajan  and
      Federmann, Christian  and
      Haddow, Barry  and
      Hokamp, Chris  and
      Huck, Matthias  and
      Logacheva, Varvara  and
      Pecina, Pavel",
    booktitle = "Proceedings of the Tenth Workshop on Statistical Machine Translation",
    month = sep,
    year = "2015",
    address = "Lisbon, Portugal",
    publisher = "Association for Computational Linguistics",
    url = "https://aclanthology.org/W15-3049/",
    doi = "10.18653/v1/W15-3049",
    pages = "392--395"
}

@article{gemini2023,
  title={Gemini: A Family of Highly Capable Multimodal Models},
  author={Gemini Team, Google},
  journal={arXiv preprint arXiv:2312.11805},
  year={2023},
  url={https://arxiv.org/abs/2312.11805}
}

@misc{github2022copilot,
  title={{GitHub Copilot is generally available to all developers}},
  author={{GitHub}},
  year={2022},
  month={June},
  howpublished={\url{https://github.blog/2022-06-21-github-copilot-is-generally-available-to-all-developers/}},
  note={Accessed: 2025-09-24}
}

@misc{google2023duetai,
  title={{Duet AI in Google Cloud is now generally available}},
  author={{Google Cloud}},
  year={2023},
  month={May},
  howpublished={\url{https://cloud.google.com/blog/products/application-modernization/introducing-duet-ai-for-google-cloud}},
  note={Accessed: 2025-09-24]}
}

@misc{aws2023codewhisperer,
  title={{Amazon CodeWhisperer is now generally available}},
  author={{Amazon Web Services}},
  year={2023},
  month={April},
  howpublished={\url{https://aws.amazon.com/blogs/aws/amazon-codewhisperer-free-for-individual-use-is-now-generally-available/}},
  note={Accessed: 2025-09-24}
}

@misc{windsurf,
  author={{Windsurf}},
  title={{Code suggestions powered by everything you've done}},
  howpublished={\url{https://windsurf.com/tab}},
  note={n.d.},
  urldate={2025-09-24}
}

@misc{cursor,
  author = {{Anysphere Inc.}},
  title = {{Cursor: The AI-first Code Editor}},
  howpublished = {\url{https://cursor.sh/}},
  note = {n.d.},
  urldate = {2025-09-24}
}
